\documentclass[lettersize,journal]{IEEEtran}
\usepackage{amsmath,amsfonts}
\usepackage{algorithmic}
\usepackage{algorithm}
\usepackage{array}
\usepackage[caption=false,font=normalsize,labelfont=sf,textfont=sf]{subfig}
\usepackage{textcomp}
\usepackage{stfloats}
\usepackage{url}
\usepackage{verbatim}
\usepackage{graphicx}
\usepackage{cite}

\usepackage{soul}
\usepackage{listings}
\usepackage{multirow}
\usepackage{colortbl}
\usepackage{hyperref}
\usepackage{stmaryrd}
\usepackage{inconsolata}
\usepackage{threeparttable}

\hypersetup{
colorlinks=true,
linkcolor=black,
citecolor=black,
urlcolor=black,
}
\definecolor{Lavender}{RGB}{240, 240, 240}
\definecolor{darkgray}{RGB}{150, 150, 150}
\definecolor{BlueViolet}{RGB}{138, 43, 226}
\definecolor{SeaGreen}{RGB}{60, 179, 113}
\usepackage{booktabs}
\usepackage{soul}

\newcounter{reviseCounter}

\begin{document}
\title{LuaTaint: A Static Analysis System for Web Configuration Interface Vulnerability of Internet of Things Devices}

\author{Jiahui Xiang,~\IEEEmembership{Student Member,~IEEE}, Lirong Fu, Tong Ye, Peiyu Liu,~\IEEEmembership{Member,~IEEE}, \\ Huan Le, Liming Zhu, Wenhai Wang
\thanks{
\emph{(Corresponding author: Peiyu Liu; Wenhai Wang.)}

J. Xiang, T. Ye, P. Liu, and W. Wang are with the College of Control Science and Engineering, Zhejiang University, Hangzhou 310027, China (e-mail: xiangjh@zju.edu.cn; tongye@zju.edu.cn; liupeiyu@zju.edu.cn; zdzzlab@zju.edu.cn).

L. Fu is with the School of Cyberspace Security, Hangzhou Dianzi University, Hangzhou 310012, China (e-mail: fulirong007@zju.edu.cn).

H. Le and L. Zhu are with the China Tobacco Zhejiang Industrial CO.,LTD, Hangzhou 310009, China (e-mail: lehuan@zjtobacco.com; zlm@zjtobacco.com).

Copyright (c) 2024 IEEE. Personal use of this material is permitted. However, permission to use this material for any other purposes must be obtained from the IEEE by sending a request to pubs-permissions@ieee.org.}
}

\markboth{XIANG \MakeLowercase{\textit{et al.}}: LuaTaint: A Static Analysis System for Web Configuration Interface Vulnerability of I\MakeLowercase{o}T Devices}%
{XIANG \MakeLowercase{\textit{et al.}}: LuaTaint: A Static Analysis System for Web Configuration Interface Vulnerability of I\MakeLowercase{o}T Devices}


\maketitle
\begin{abstract}
The diversity of web configuration interfaces for IoT devices has exacerbated issues such as inadequate permission controls and insecure interfaces, resulting in various vulnerabilities. Owing to the varying interface configurations across various devices, the existing methods are inadequate for identifying these vulnerabilities precisely and comprehensively. This study addresses these issues by introducing an automated vulnerability detection system, called LuaTaint. It is designed for the commonly used web configuration interface of IoT devices. 
LuaTaint combines static taint analysis with a large language model (LLM) to achieve widespread and high-precision detection. The extensive traversal of the static analysis ensures the comprehensiveness of the detection. The system also incorporates rules related to page handler control logic within the taint detection process to enhance its precision and extensibility. Moreover, we leverage the prodigious abilities of LLM for code analysis tasks. By utilizing LLM in the process of pruning false alarms, the precision of LuaTaint is enhanced while significantly reducing its dependence on manual analysis. We develop a prototype of LuaTaint and evaluate it using 2,447 IoT firmware samples from 11 renowned vendors. LuaTaint has discovered 111 vulnerabilities. Moreover, LuaTaint exhibits a vulnerability detection precision rate of up to 89.29\%.
\end{abstract}


\section{Introduction}
\IEEEPARstart{T}{he} Internet of Things (IoT) is a rapidly growing industry, and IoT devices are used in many areas closely related to human life and productivity \cite{whitmore2015internet}. Numerous IoT devices with various system architectures lack traditional security mechanisms, resulting in various security risks in these IoT devices \cite{ALABA201710, meneghello2019iot}. In recent years, firmware web vulnerabilities have been frequently reported, and IoT devices have become popular targets for malicious attacks \cite{Threat2020Report, Common2022Attacks, Millions2021Routers, khandelwal2018thousands, liu2023iot}. As the web interface becomes the primary management interface for IoT devices, the firmware of these devices often embeds web servers that run web applications to communicate with IoT users.
The web application embedded in the device is called the web configuration interface, which is generally used for user interaction and web interface configuration.

Given the security concerns surrounding IoT devices and web configuration interfaces, we conducted a statistical analysis of the firmware web vulnerabilities. Our results reveal a significant increase in the total number of vulnerabilities associated with web configuration interfaces in recent years. The notable proportion of taint-based vulnerabilities is particularly noteworthy, in which code injection and remote code execution vulnerabilities accounted for approximately 32\%. Taint-based vulnerabilities typically arise from the inadequate validation or handling of user-provided data by an application during the input processing. This can lead to the injection of malicious input or code into the program, thereby creating security vulnerabilities during execution. Therefore, it is of utmost importance to develop an precise and comprehensive automated detection method to detect these vulnerabilities.

Many techniques exist for identifying firmware vulnerabilities at the level of web services. For example, SATC \cite{chen2021sharing} uses shared front-end and back-end keywords to start static taint analysis and identify security vulnerabilities. SRFuzzer \cite{zhang2019srfuzzer} deploys an automated fuzzy testing framework to analyze the web server module of small office and home office routers. However, both of them do not primarily target the web configuration interface of the firmware but rather focus on back-end binaries. From the perspective of a web configuration interface, certain scholars have undertaken relevant research endeavors. Costin et al. \cite{costin2016automated} propose a scalable and fully automated dynamic analysis framework to discover vulnerabilities of the web configuration interface in firmware. WMIFuzzer \cite{wang2019discovering} is designed to discover vulnerabilities in commercial off-the-shelf IoT devices by fuzzing their web configuration interfaces.
Most recent studies have relied on existing web analysis tools and used dynamic analysis to simulate the firmware \cite{costin2016automated, chen2016towards, wang2019discovering, srivastava2019firmfuzz, yin2021firmhunter}. As is well known, dynamic analysis commonly encounters issues such as limited coverage, significant resource overhead, and reliance on specific environments.

Currently, existing methods cannot effectively detect vulnerabilities within firmware web configuration interfaces on a large scale. Current static analysis tools have difficulty addressing the intricate IoT web configuration interface. To fill the research gap, we endeavor to develop a static analysis system for taint-based vulnerabilities centered on web interfaces in IoT devices. We take Lua Configuration Interface (LuCI), a widely utilized web configuration interface in firmware, as our focal point. The vulnerability detection tool that we designed for the firmware web configuration interface faces three key challenges in real IoT devices:
\begin{itemize}
\item[$\bullet$] The variable classes and complex data structures of the development language Lua commonly used in web configuration interfaces increase the uncertainty of the code during static analysis.
\item[$\bullet$] The intricate dispatching rules of web configuration interfaces pose challenges to implementing vulnerability detection systems in IoT devices. 
\item[$\bullet$] Static taint analysis often struggles with overtaint and undertaint issues, making it difficult to determine an effective balance between soundness and completeness.
\end{itemize}

In this study, we address the aforementioned challenges by developing an automated system to detect common vulnerabilities in firmware web configuration interfaces. (a) This system leverages flow-sensitive, context-sensitive, and field-sensitive static taint analyses to detect potential vulnerabilities precisely. It effectively addresses the uncertainties arising from the characteristics of Lua. (b) We incorporate the dispatching rules from the web configuration interface in IoT devices into the framework-adapted module, thereby improving the adaptability of the system to the current task. (c) Considering the excellent performance of large language model (LLM) in code analysis, we utilized LLM for false alarm pruning to further enhance detection. The combination of the static analysis with the dispatching rules and the support of LLM effectively ensures the soundness and completeness of detection.

We develope a specialized system called LuaTaint to refine our bug detection process. We transform the source code into an abstract syntax tree (AST) to identify sensitive data and conduct control flow analysis. To gain a more nuanced understanding of the behavior of the program, we employ data flow analysis to help establish data flow constraints for nodes within the control flow graph (CFG). Subsequently, we use a customized taint analysis approach to identify their vulnerabilities and document details. We efficiently sift through false alarms using a strategic questioning design for LLM. Ultimately, we craft the proof of concepts manually to inspect the flagged alarms and confirm their vulnerability status further. Specifically, we use Firmware Analysis Plus \cite{firmware-analysis-plus}, a QEMU-based \cite{qemu} dynamic firmware analysis platform, to simulate the running firmware system. By interacting with the web interface of the emulated firmware, we introduce constructed malicious payloads (mainly URLs and JSON-formatted form data) to exploit potential vulnerabilities. If the system executes a malicious command, we classify the alarm as a true bug.

Our system, LuaTaint, comprises four key components: a parsing and control flow analyzer, a reaching definitions analyzer, a framework-adapted taint analyzer, and a pruning processor using LLM. To assess the capability of LuaTaint in pinpointing vulnerabilities within IoT devices, we deployed it on 2,447 firmware samples sourced from 11 vendors. We completed the verification of 145 firmware samples, and LuaTaint identified 111 vulnerabilities. Additionally, we tested LuaTaint precision to about 89.29\% on a dataset with 21 firmware samples. We compared LuaTaint with Semgrep, an advanced open-source static analysis tool, and found that LuaTaint obviously outperforms Semgrep in vulnerability detection for Lua and LuCI frameworks. The results indicate that LuaTaint can conduct effective and practical security analyses of IoT firmware web configuration interfaces.

The contributions of this study are as follows:

\begin{itemize}
\item[$\bullet$] We implement the first automated static detection system for web configuration interface vulnerabilities of IoT devices.
\item[$\bullet$] We employ LLM to effectively prune false alarms, demonstrating the excellent performance of LLM in evaluating program security.
\item[$\bullet$] We evaluate LuaTaint in real large-scale firmware scenarios and find 111 vulnerabilities.
\end{itemize}

To encourage future research, we made our source code publicly available.\footnote{\href{https://github.com/miko99jh/LuaTaint/blob/main/README.md}{https://github.com/miko99jh/LuaTaint/blob/main/README.md}}

The remainder of this article is organized as follows. Section \ref{Background} provides the background and motivation of this work. Next, we illustrate the design and implementation of our static analysis system, LuaTaint, in Sections \ref{LuaTaint} and \ref{Implementation}. We demonstrate the efficacy of LuaTaint through experiments and case studies on real-world firmware samples in Section \ref{Evaluation}. The limitation of LuaTaint is discussed in Section \ref{Discussion}. Section \ref{RelatedWork} briefly reviews related research from the perspective of firmware and web interface security. Finally, we present concluding remarks and future work directions of this study in Section \ref{Conclusion}.
 
\section{Background and Motivation} \label{Background}
In this section, we first provide an overview of the OpenWrt system and its web configuration interface, LuCI. Next, we conduct an observation of vulnerabilities in IoT devices to verify the necessity of the current work. Finally, we present an example of vulnerability to illustrate the motivation for this study.

\subsection{OpenWrt and LuCI}
OpenWrt is essentially a highly modular and automated embedded Linux system. It is often used in industrial control devices, telephones, miniature robots, smart homes, routers, and voice over internet protocol devices \cite{holt2018openwrt}. Currently, hardware devices enabled with Openwrt include various model versions from popular brands, such as 8devices, ASUS, D-Link, GL.iNet, Huawei, and TP-Link \cite{openwrtstart2023}.

The LuCI serves as a unified configuration framework based on Lua, which is employed to manage web configuration interfaces of OpenWrt\cite{openwrtluci2021}. Lua is a lightweight, powerful, and efficient programming language that is primarily designed to embed within applications \cite{lua2023}. LuCI is structured using the model-view-controller (MVC) architecture pattern \cite{leff2001web}, which facilitates efficient management of the various components of the interface. This framework is notable for its significant reusability and extensibility; however, during the development phase, there is an increased risk of web vulnerabilities. A notable concern is the remote injection vulnerability, which allows an attacker to manipulate the back-end system remotely by injecting system commands or code directly into the server. These vulnerabilities often emerge when the application is designed to provide specific remote command interfaces, particularly when developers ignore to properly check user input parameters or when these checks are insufficiently rigorous. They are always evident in the web configuration interfaces of routers, firewalls, intrusion detection systems, and similar devices.

\begin{figure}
\setlength{\abovecaptionskip}{-1mm}
\includegraphics[width=0.97\linewidth]{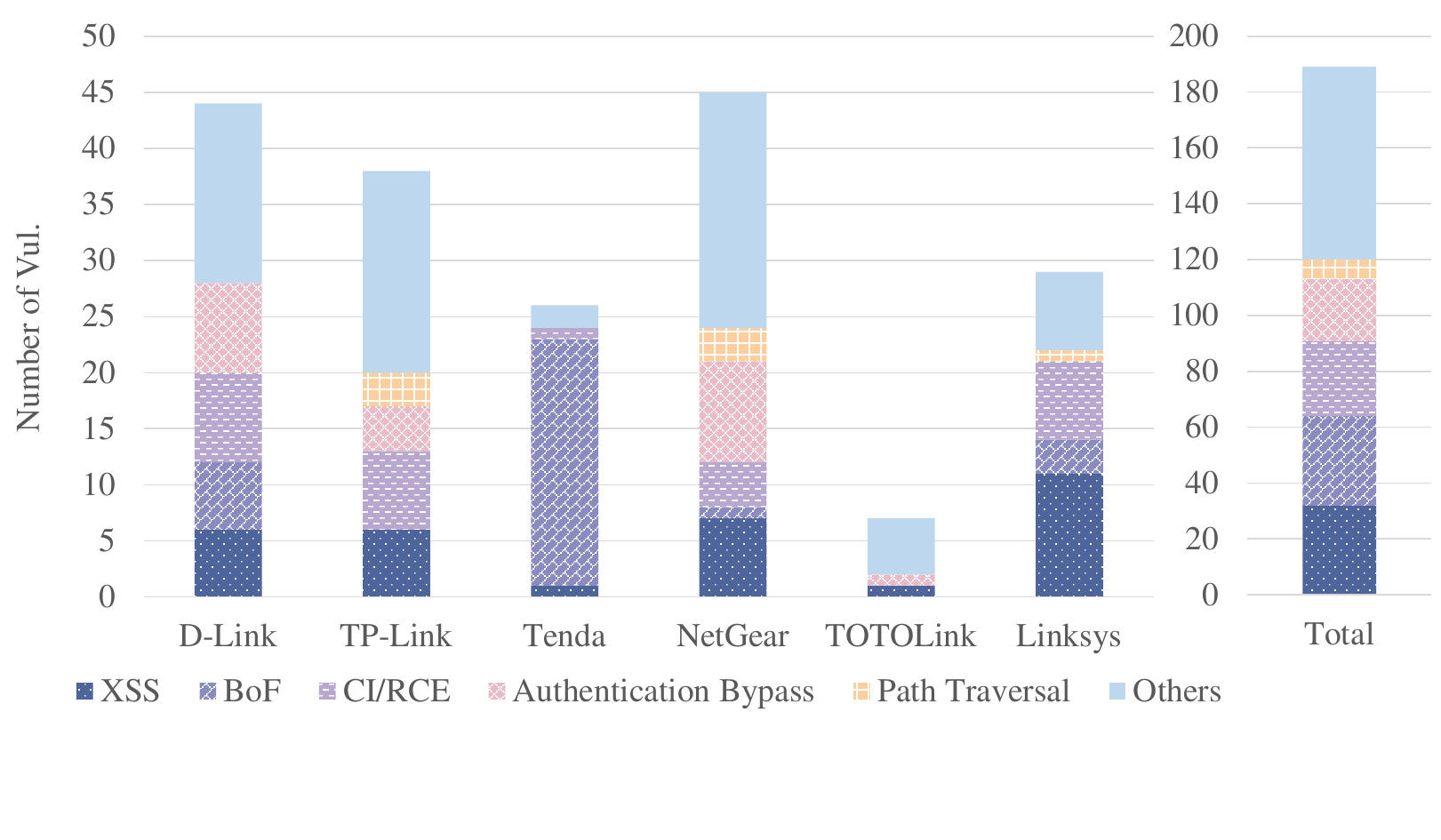}
\centering
\caption{\label{fig:1}Statistics on the number of firmware web vulnerabilities listed by type.}
\end{figure}

\subsection{Observation}
We conducted a statistical survey to obtain an intuitive sense of the web vulnerabilities that exist in firmware. By collecting firmware web vulnerabilities reported on the Common Vulnerabilities and Exposures (CVE) platform, we randomly selected 189 firmware web-related vulnerabilities from six vendors, including D-Link, TP-Link, Tenda, NetGear, TOTOLink, and Linksys, for sampling statistics. Figure \ref{fig:1} shows that in firmware web vulnerabilities, cross-site scripting (XSS) vulnerabilities and buffer overflow (BoF) vulnerabilities account for a relatively large proportion of the statistics, both accounting for 16.9\%; command injection (CI) vulnerabilities and code execution (RCE) vulnerabilities are also very typical, accounting for approximately 14.3\%. As observed, taint-type vulnerabilities (CI, RCE, XSS, and path traversal) constitute the majority of the web vulnerabilities in IoT devices.

\begin{figure}
\setlength{\abovecaptionskip}{-1mm}
\includegraphics[width=0.6\linewidth]{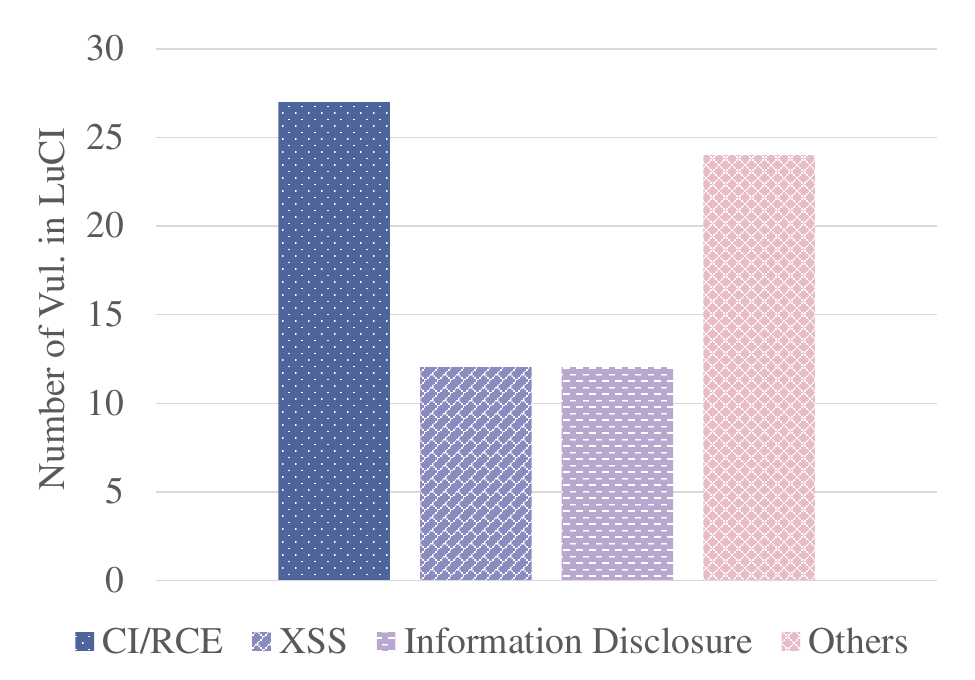}
\centering
\caption{\label{fig:2}Statistics based on the number of vulnerability types associated with the LuCI.}
\end{figure}

We also performed an analysis of the CVE vulnerabilities associated with LuCI and observed 75 related vulnerabilities. The reports of vulnerabilities have increased significantly since around 2018. As shown in Figure \ref{fig:2}, when categorizing these vulnerabilities, it is evident that CI/RCE, XSS, and information leakage vulnerabilities are the most common, constituting approximately 36\%, 12\%, and 12\% of the total vulnerabilities, respectively. This highlights the wide range of security threats that currently affect LuCI.

As evident from the aforementioned two studies, taint-based vulnerabilities have emerged as the most prevalent. Taint-based vulnerabilities arise when data flows through a software component without proper validation or sanitization. Common instances of these vulnerabilities include CI, RCE, XSS, Structured Query Language (SQL) injection, and path traversal. Furthermore, insights derived from Open Web Application Security Project (OWASP) TOP10 highlight the significant relevance and prevalent exploitation of taint-based vulnerabilities in real-world attacks \cite{owasp2021topten}. Therefore, our research focuses on the detection of taint-based vulnerabilities.

\begin{figure*}
\setlength{\abovecaptionskip}{0mm}
\includegraphics[width=0.8\linewidth]{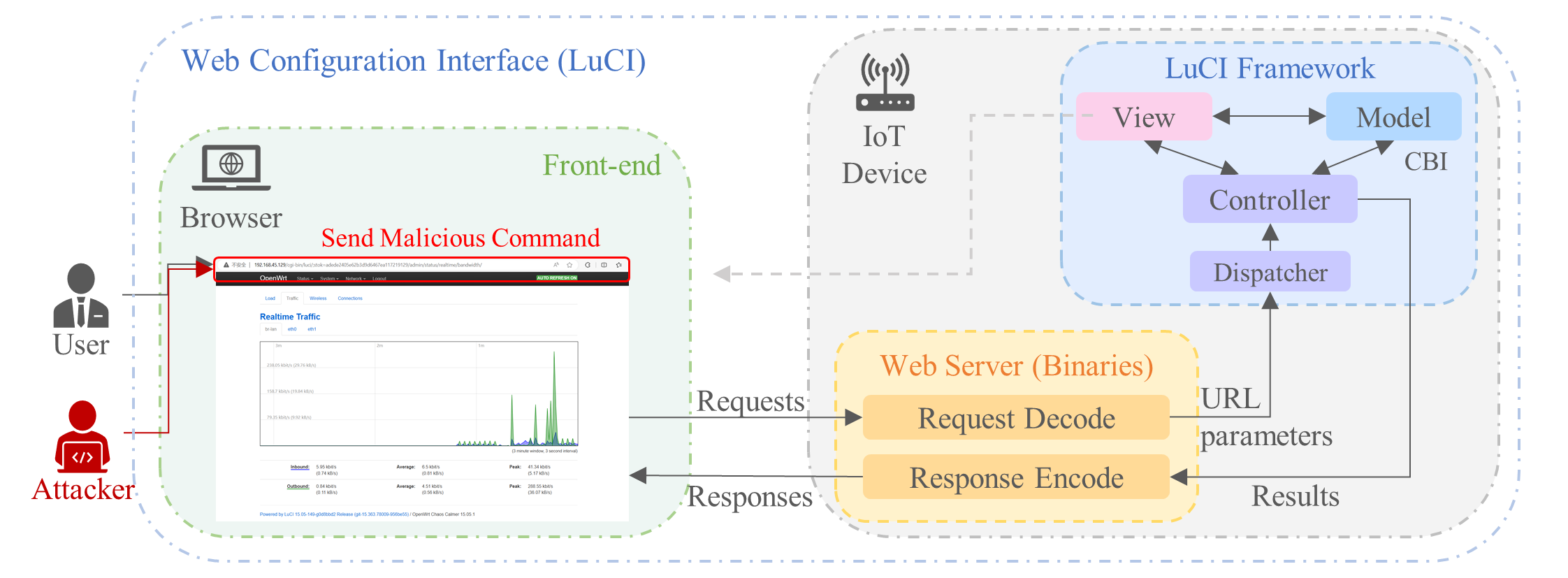}
\centering
\caption{\label{fig:3} Example of a command injection vulnerability in LuCI. After a user executes an operation in the left browser’s front-end interface, the web server receives the HTTP request and decodes the message. This message then proceeds to the LuCI framework for further processing. Once processed, LuCI encodes the results into a response message, which is then sent back to the front-end interface.}
\end{figure*}

\subsection{Motivation}
Typically, IoT devices feature a web interface for configuring the system or interacting with the external environment. The web server processes Hypertext Transfer Protocol (HTTP) requests from the front-end and then invokes the back-end binary for further processing. An attacker can craft malicious inputs targeting the front-end, potentially compromising the integrity of the corresponding back-end binary.

\textbf{Vulnerability Example:} Figure \ref{fig:3} shows an example of how the above process is executed in an OpenWrt system. Using the web configuration interfaces \texttt{"admin/status/real-time/bandwidth\_status"} and \texttt{"admin/status/realtime/wireless\_status"} via a Uniform Resource Locator (URL), an end user can access the real-time bandwidth status or the wireless network status of the current system. Unfortunately, when the user is applying the current interfaces, these two endpoints correspond to the functions \texttt{action\_bandwidth} and \texttt{action\_wireless}, which do not perform any cleanup checks when generating executable commands for logging traffic. An attacker can add additional commands to the URL to execute arbitrary commands. For example, if an attacker sends a malicious reboot command directly to the back-end via the URL \texttt{"http://IP:Port/cgi-bin/luci/admin/status/realtime\\/bandwidth\_status/eth0\$(reboot)"}, the device is remotely attacked. This command injection vulnerability (CVE-2019-12272) exists in OpenWrt LuCI 0.10 and earlier versions.

Currently, there is a significant gap in the detection methods that can effectively identify taint-based vulnerabilities in Lua-based web configuration interfaces. This study aims to address this gap by developing an automated vulnerability detection system using LuCI as the target.

\section{LuaTaint} \label{LuaTaint}
In this section, we outline the core principles of LuaTaint, starting with a comprehensive overview. We then investigate the following four essential elements of our automated vulnerability detection system: the parsing and control flow analyzer, the reaching definitions analyzer, the framework-adapted taint analyzer, and the pruning processor with LLM.
 
\subsection{Overview of LuaTaint}
\begin{figure}
\setlength{\abovecaptionskip}{-1mm}
\includegraphics[width=1.0\linewidth]{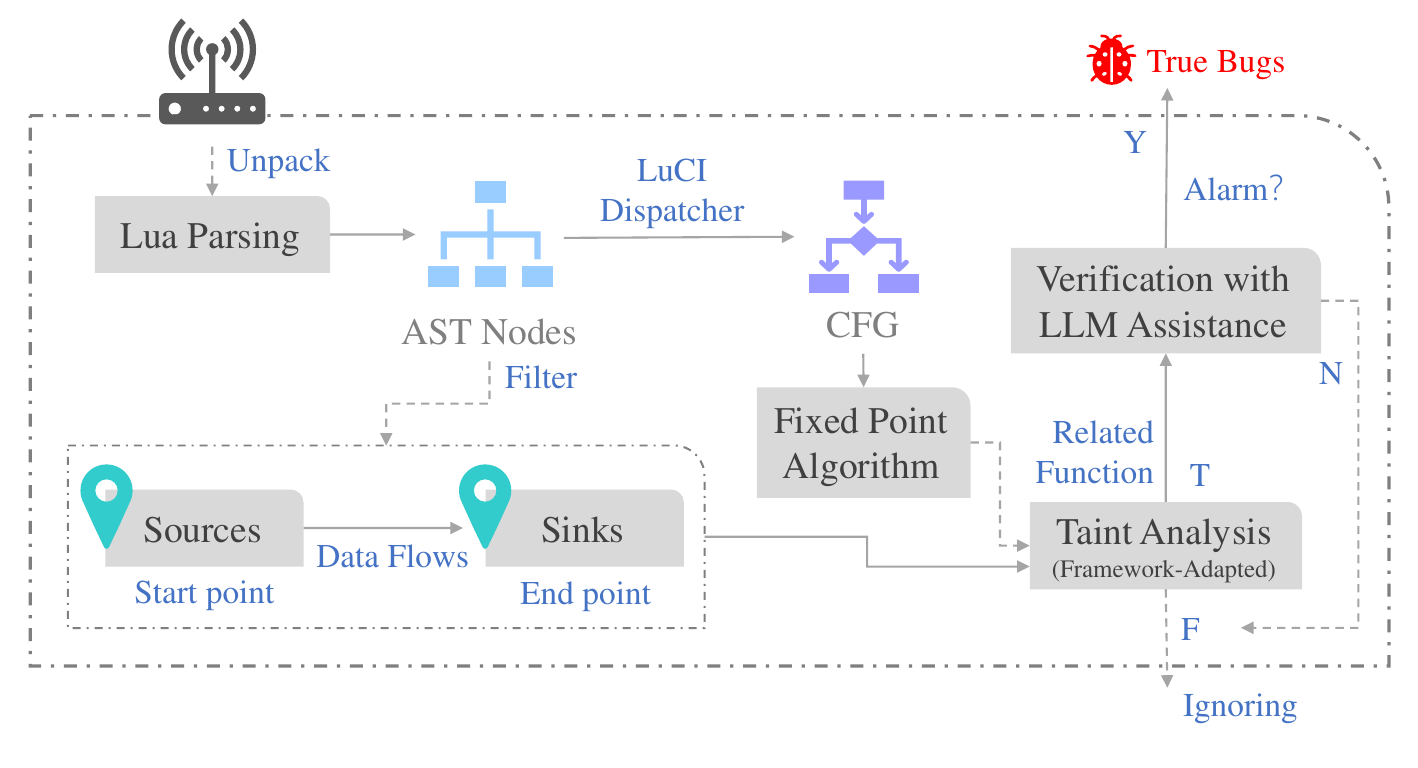}
\centering
\caption{\label{fig:4}Architecture of LuaTaint. LuaTaint analyzes the control flow and data flow of the LuCI framework, formulates rules based on the characteristics of the web configuration interface, and employs static taint analysis and LLM to automate the detection of web vulnerabilities.}
\end{figure}

Figure \ref{fig:4} provides an overview of LuaTaint, which takes the firmware image as input and ultimately reports potential bugs in IoT firmware by the following procedures. 

First, we unpack the firmware image using an existing firmware unpacker (i.e., binwalk \cite{heffner2014binwalk}). LuaTaint identifies the web configuration interface within the extracted files according to their type. Typically, the LuCI framework is located at \texttt{"/user/lib/lua/luci"} in the root directory, with its core framework coded in Lua, apart from the front-end web pages, which are built with HTML templates.

In order to conduct a structured analysis of the source code, LuaTaint first converts the Lua code within the LuCI framework into an AST and then into a CFG. To ensure a thorough analysis, the CFG is then enriched with an interprocess analysis that reflects the specific dispatching rules of LuCI and includes essential functions that are not directly called. To analyze the path of data flow in the program, the system conducts data flow analysis on the CFG to identify constraints at CFG nodes using a fixed-point algorithm alongside reaching definitions analysis. LuaTaint uses AST to pinpoint critical components, such as sources, sinks, and sanitizers, within the LuCI framework. It begins its analysis by identifying sinks from a reverse engineering perspective and then tracing the data flow into sensitive parameters based on a field-sensitive algorithm to detect potentially dangerous external inputs. This approach to input sensitive static taint analysis, grounded in AST, helps to locate potential vulnerabilities accurately. Additionally, LuaTaint utilizes LLM to prune false alarms, thereby improving the precision of its vulnerability detection.

The four components in our system correspond to four analysis processes: parsing and control flow analysis, reaching definitions analysis, framework-adapted taint analysis, and pruning with LLM assistance. They are shown below.

\setlength{\parskip}{0.1cm plus1mm minus1mm}

\subsection{Parsing and Control Flow Analysis}
In order to accurately identify the syntactical structure in the program in the static analysis, the LuCI framework source code within the IoT firmware requires conversion to AST by syntactical analysis. Each node in the AST is representative of a unique structural element in the source code, facilitating the precise delineation of relationships such as function invocations, assignment statements, and operational directives. In comparison with intermediate representations, the AST maintains fidelity closer to the original syntactic structure and is considered to be more conducive to rapid type validation.

In our practical work, several syntax errors were observed during the parsing of the source code. These issues primarily arose from the unconventional syntax formats inherent in firmware (more specifically, the unconventional escape characters). The strict requirements of the parser for handling escape characters hindered its ability to parse certain expressions containing these characters. This obstructed the generation of the AST and further analysis. To address this challenge, we meticulously compiled a list of unparseable syntax formats. Before parsing the source code, we conducted a thorough pre-scan of the entire codebase and used regular expressions to replace problematic syntax formats. This method facilitated seamless parsing. For example, the character \texttt{"\textbackslash*"} in the expression \texttt{val:match("\textbackslash\^{}[0-9\textbackslash*\#!\textbackslash\%.]+\textbackslash\$")} and the character \texttt{"\textbackslash-"} in the expression \texttt{string.gsub(value, "\^{}\textbackslash-", "\textbackslash\textbackslash-")} cannot be fully recognized by parser. We will replace them with the more canonical escape characters \texttt{"\%*"} and \texttt{"\%-"} to ensure proper parsing and functionality.

AST typically varies based on the programming language and often lacks information on control flow. Therefore, it necessitates conducting a separate control flow analysis based on the AST. To construct a CFG from an AST, we perform a traverse of the AST to establish and assign predecessor and successor relationships between nodes. These predecessor and successor attributes refer to other nodes in the graph. Each CFG is typified by the presence of an initial entry node, which uniquely engenders a successor, and a terminal exit node, which exclusively accommodates a precursor. The interconnection of intermediate nodes is achieved through precursor and successor relationships, resulting in a CFG represented as a list of nodes comprising CFG nodes. This procedure involves the definition of distinct traversal protocols for various node typologies, thus facilitating the establishment of connective relationships between CFG nodes during the evaluation of Lua statements and expressions.

When constructing a CFG for Lua using AST nodes, it's crucial to grasp Lua's control structures thoroughly. Lua provides a regular collection of control structures, such as \texttt{if} for conditional judgment and \texttt{while}, \texttt{repeat}, and \texttt{for}, for iteration \cite{lua2023}. Each control structure statement has a terminator: the \texttt{end} is used for \texttt{if}, \texttt{while}, and \texttt{for}, and the \texttt{until} is used for \texttt{repeat}. Notably, numerous variants of the Lua syntax structures exist in various forms. For example: 
(a) There are two types of \texttt{for} statements: numeric for loops and generalized for loops;
(b) No statements inside the body of \texttt{while}, \texttt{for}, \texttt{repeat}, and \texttt{if} control structures should be allowed;
(c) Lua functions can accept a variable number of arguments, similar to C using \texttt{(...)} in the function argument list. Current control flow analysis methods struggle to adequately handle Lua's control structures. Thus, when establishing connections between CFG nodes, we consider all Lua syntax structures, including those mentioned above, to ensure the feasibility of control flow analysis.
 
The analysis of the control flow also includes examining the relationships between function calls. In this context, we use multivariate interprocedural analysis \cite{schwartzbach2008lecture} to address function calls. Each analysis within a single function is linked to a specific abstract domain. When a function is called, we treat it as a new separate analysis, incorporating the CFG of the called function into the overarching CFG. 
To maintain correct semantics across various function scopes, we employ shadow variables \cite{micheelsen2016static}. For context-sensitive analysis, distinguishing calls to the same function from various locations is crucial. This involves creating a separate instance of the called function for every location from which it is called, thereby preventing any mix-up of interactions between calls from various locations.

\textbf{LuCI Dispatcher.} To enhance the performance of the analysis for the given task, we also need to consider the characteristics of the analysis object. An entry function is dedicated to the display of web pages in the controller of LuCI. Considering the definition in the dispatcher, the prototype of the controller's function for generating URLs is \texttt{entry(path, target, title=nil, order=nil)}, where \texttt{path} is the path to access, which is given as an array of strings. For example, if the path is written as \texttt{"\{"admin", "loogson", "control"\}"}, you can access this script in your browser by assessing to \texttt{"http://192.168.1.1/cgi-bin/luci/admin/loogson/co-\\ntrol"}. The control flow analysis cannot discern this type of call relationship as an explicit call, so it is necessary to utilize the dispatching rules of the LuCI framework to set all called function nodes to the CFG list.

\textbf{Efficiency Optimization.} We take the LuCI framework as system input and generate a dictionary of function definitions by traversing the nodes in the CFG. This dictionary compiles all the function definitions within LuCI, which is crucial for conducting taint analysis and identifying vulnerabilities across various files. Inevitably, querying and tracking the called function consumes considerable computational resources, particularly for large projects. To enhance efficiency, we approximate in the dictionary using function node names as keys instead of the nodes while keeping the nodes unchanged as values. When we encounter a function node with an existing name in a new location, the new node replaces the old one. Our real-world experiments demonstrate that this simplification does not result in false negatives for our current tasks, as it does not lead to any loss of functional-level information within the current file.

\subsection{Reaching Definitions Analysis}
To determine how dangerous data flows move along the execution path of the program and where sensitive data are ultimately transferred, we perform a data flow analysis based on the CFG described above. Here, we use the \textbf{reaching definitions analysis} to track data flow information \cite{schwartzbach2008lecture}.

A simple way to conduct reaching definitions analysis on a program is to set up data flow equations for each node of the CFG and iteratively compute the output of the local input of each node until the entire system state is stable. We use data flow constraint equations to describe the constraint relationships between each node of the CFG. Data flow constraints are denoted by $\llbracket v_i \rrbracket$, which associates the values of a node with its neighbors. For a CFG containing nodes $V=\{v_1,v_2,\ldots,v_n\}$, we analyze it on the lattice $L^n$. The lattice used for this analysis is a power set lattice of all the assignment nodes in the program. Suppose that the relationship between a node $v_i$ and its neighboring nodes is expressed as the data flow equation: $\llbracket v_i\rrbracket=F_i(\llbracket v_1\rrbracket,\ldots,\llbracket v_n\rrbracket)$, then the equations for all nodes can be combined into a single function $F:L^n\xrightarrow{}L^n$ as a constraint system.

The data flow constraint system performs the analysis in reverse by connecting the constraints of all the predecessor nodes of all nodes. This can be expressed as the following function:
\begin{equation}
    JOIN(v)=\bigcup_{w\in pred(v)}\llbracket w\rrbracket
\end{equation}

For the function $F:L^n\xrightarrow{}L^n$ in each iteration, the update rules for the data flow constraints are categorized into two types for two types of nodes: assignment and non-assignment nodes.

For the assignment node, because the assignment node changes the value of the current variable, it is necessary to discard the assignment value of the variable corresponding to the current node and replace it with the current assignment node:
\begin{equation}\label{eq:eq4}
    \llbracket v\rrbracket=JOIN(v)\downarrow id\cup\{v\}
\end{equation}

For all non-assigned nodes, just join all constraints before them:
\begin{equation}
    \llbracket v\rrbracket=JOIN(v)
\end{equation}

The $\downarrow$ function here removes from the $JOIN$ result all assignments to the variable \emph{id}.

Based on the \textbf{Fixed-Point Theorem} \cite{schwartzbach2008lecture}, in a lattice $L$ with finite height, every monotone function $f$ has a unique least fixed point. The above data flow constraint system is computed by iteratively solving the system until no further change occurs between two adjacent iterations of the computation, that is, the \textbf{fixed point} is reached. This point is also the result of the propagation of all the assignments through the program, and this information can be used to infer the presence of any possible dangerous data flows.

\textbf{Algorithm \ref{alg:alg1}} details the process of reaching definitions analysis. To reduce the complexity of the algorithm, we enhance the iterative algorithm by implementing the \textbf{Worklist Algorithm} using a worklist, denoted by q, to store all the nodes within a CFG. The process continues provided that the worklist \emph{q} remains nonempty, iterating through the following steps until a stable state is reached. We start by updating the first node in \emph{q} based on its type, creating a new constraint \emph{new}. If this new constraint differs from the previous one, this difference indicates a change in the constraint information of the node, which must then be communicated to its successors. We iterate through each successor of this node by adding them to the worklist \emph{q} and updating their constraint information to match the new constraint. After finishing these updates in the inner loop, the first node \emph{q}[0] is removed from the worklist. In this way, we use the iterative algorithm to solve the data flow constraint relationship between program nodes.

\begin{algorithm}\small
\caption{Reaching Definitions Analysis Worklist}\label{alg:alg1}
\begin{algorithmic}
\STATE
\STATE\textbf{Input:} $A\ list\ of\ multiple\ CFG\ \boldsymbol{list_{CFG}},\ representing$\\ $all\ CFG\ structures\ contained\ in\ a\ lua\ file$. \\
\STATE\textbf{Output:} $The\ constraint\ table\ of\ each\ CFG\ \boldsymbol{CT}$ \\
\STATE\,\ 1: $CT\, =\, dict()$ \\
\STATE\,\ 2: \textbf{for} $each\, CFG$ \textbf{in} $list_{CFG}$ \textbf{do} \\
\STATE\,\ 3: \quad $q=Lattice(CFG.nodes)$ \\
\STATE\,\ 4: \quad\textbf{while} $q! = [\ ]$ \textbf{do} \\
\STATE\,\ 5: \qquad $old = CT[q[0]]$ \\
\,\ \textbf{Fixed-Point Algorithm:} \\
\STATE\,\ 6: \qquad\textbf{if} $isinstance(q[0], AssignmentNode)$ \textbf{then}\\
\STATE\,\ 7. \qquad\quad $CT[q[0]]=JOIN(q[0])\downarrow id\cup{q[0]}$ \\
\STATE\,\ 8: \qquad\textbf{else} $CT[q[0]]=JOIN(q[0])$\\
\STATE\,\ 9: \qquad\textbf{end if} \\
\STATE10: \qquad $new = CT[q[0]]$ \\
\STATE11: \qquad\textbf{if} $new! = old$ \textbf{then}\\
\STATE12: \qquad\quad \textbf{for} $each\, node$ \textbf{in} $q[0].outgoing$ \textbf{do} \\
\STATE13: \qquad\qquad $q.append(node)$ \\
\STATE14: \qquad\qquad $CT[q[0]] = new$ \\
\STATE15: \qquad\quad\textbf{end for} \\
\STATE16: \qquad\textbf{end if} \\
\STATE17: \qquad $q = q[1:]$ \\
\STATE18: \quad\textbf{end while} \\
\STATE19: \textbf{end for} \\
\end{algorithmic}
\label{alg1}
\end{algorithm}

\subsection{Framework-Adapted Taint Analysis}
To track the data flow of tainted information within the LuCI environment, we performed a framework-adapted taint analysis. 

Taint analysis can be abstracted into a ternary $\langle$\emph{sources, sinks, sanitizers}$\rangle$. Vulnerability arises only when both sources and sinks are present, and the data flow from sources to sinks occurs without passing through sanitizers. Our primary objective is to identify vulnerabilities within the LuCI framework of the firmware, necessitating a thorough understanding of the web configuration interfaces in the LuCI framework and the corresponding code logic of their invocation. We adopt a reverse perspective in our vulnerability detection approach. Initially, we focus on identifying sensitive functions within the code responsible for system operations. Subsequently, we conduct a global search to determine which interfaces invoke the identified application programming interfaces (APIs). Based on AST node retrieval, we pinpoint the specific code segments for taint analysis and determine the potential injection of external hazardous data. In contrast to a forward analysis of all web configuration interfaces, this approach is significantly efficient because it involves traversing only the sensitive functions.

Next, we illustrate the process of identifying sources, sinks, and sanitizers and conducting taint propagation analysis with field-sensitive analysis and framework-adapted module. 

\emph{Sinks:} We conduct a comprehensive search across the entire AST for nodes representing calls to sensitive functions to identify potential tainted sinks. Once we pinpoint these sensitive functions, we perform an argument-backtracking analysis to investigate the arguments used within these functions, assessing whether they contain dangerous variables fed by an external input. If the source of the data is constant or does not depend on user input, it is considered safe. However, if it is influenced by the user input, we perform a taint propagation analysis to trace how the variables are transmitted.

\emph{Sources:} In the context of the LuCI framework, we also need to determine the tainted source of the user's input. Despite the encapsulation of the HTTP protocol within the LuCI framework, it is essential to recognize the interfaces utilized by HTTP requests to determine the source of the tainted vulnerability. For example, functions such as \texttt{luci.http.formvalue} are potential entry points for untrusted data.

\emph{Sanitizers:} Properly defining the sanitizer function can minimize false positives during the detection process. In the LuCI framework, \texttt{luci.util.shellquote} is an often used sanitizer function that secures the values quoted for shell command execution. 

A complete list of sinks, sources, and sanitizers is available at here. \footnote{\href{https://github.com/miko99jh/LuaTaint/blob/main/vulnerability_definitions/all_trigger_words.pyt}{Trigger-words List.}}

\textbf{Field-Sensitive Analysis.} 
Lua's dynamic typing challenges static analysis by obscuring variable types until runtime. Additionally, its reliance on tables for complex data types complicates the tracking of their structures. To enhance precision in pinpointing vulnerabilities and reduce false positives, we integrate field-sensitive analysis into our taint tracking process and propose a field-sensitive taint propagation algorithm for Lua. This method involves mapping each key in a table to its corresponding value and maintaining this association throughout the analysis. For instance, in Listing \ref{lst:2}, a nonfield-sensitive approach might fail to differentiate between the uses of \texttt{bean.name} and \texttt{bean.gender} because it does not consider the attribute level of the table. This approach can incorrectly treat the data retrieved from the \texttt{Bean} object as equally untrustworthy. In contrast, field-sensitive analysis correctly identifies that \texttt{bean.name} originates from an untrusted environmental variable, whereas \texttt{bean.gender} is a secure value. Consequently, it accurately flags \texttt{command1} as potentially hazardous instead of \texttt{command2}. Our analysis distinguishes between table variable attributes originating from untrusted sources and those from trusted inputs. 

\lstset{
 basicstyle=\ttfamily\fontsize{8}{9}\selectfont,
 columns=fixed,
 numbers=left,                                        
 numberstyle=\tiny\color{gray},                       
 xleftmargin=1em,                                     
 aboveskip=6pt, 
 backgroundcolor=\color[RGB]{245,245,244},            
 keywordstyle=\color{blue},                 
 morekeywords={local,end,function,return},
 numberstyle=\footnotesize\color{darkgray},
 emph={os,execute,getenv,setmetatable},
 emphstyle=\color{magenta},
 moredelim=[is][\color{cyan}]{\$}{\$},
 moredelim=[is][\color{BlueViolet}]{\@}{\@},
 moredelim=[is][\color{SeaGreen}]{\&}{\&},
 showstringspaces=false,                              
 label=lst:2,
}
\begin{lstlisting}[caption={A field-sensitive example for taint propagation analysis.},captionpos=b]
local Bean = {}
$Bean$.__index = $Bean$
function $Bean$:new($name$, $gender$)
    local @self@ = setmetatable({}, $Bean$)
    @self@.&name& = $name$
    @self@.&gender& = $gender$
    return @self@
end
local $name$ = os.getenv("name")
local $bean$ = $Bean$:new($name$, "male")
local $command1$ = "echo " .. $bean$.&name&
local $command2$ = "echo " .. $bean$.&gender&
os.execute($command1$)
os.execute($command2$)
\end{lstlisting}

The taint propagation algorithm for determining the taint flow is presented in \textbf{Algorithm \ref{alg:alg2}}. It also illustrates the mechanism we have implemented to address field sensitivity. The critical procedure involves monitoring the tainted variable within a collection of the identified assignment nodes. For field sensitivity, the aim is to identify the tainted attributes tied to the untrusted table variable. The algorithm then checks whether these tainted variables or attributes are present in the variables assigned to the subsequent nodes. If these variables or attributes are not found in the subsequent nodes, the algorithm removes the assignment node from the list of taint flows. The function $lat.in$ illustrates a path-reachable relationship between the two nodes. Moreover, the $trace\_tainted\_attr$ function is designed to identify the tainted attributes of a table if the current node represents a table variable that has been assigned. In this manner, tainted data flows from the sources to the sinks in the program can be located with a fine granularity.

\begin{algorithm}\small
\caption{Taint Propagation Algorithm}\label{alg:alg2}
\begin{algorithmic}
\STATE
\STATE\textbf{Input:} $The\ lattice\ of\ CFG\ nodes\ \boldsymbol{lat},\ the\ source\ node$\\ $\boldsymbol{source},\ the\ list\ of\ Assignment\, Node\ \boldsymbol{list_{AN}}$ \\
\STATE\textbf{Output:} $The\ list\ of\ node\ in\ Tainted\, Flow\ \boldsymbol{list_{TF}}$ \\
\STATE\,\ 1: $tainted\_attr=list(), list_{TF}=[source]$ \\
\STATE\,\ 2: \textbf{for} $node$ \textbf{in} $list_{AN}$ \textbf{do} \\
\STATE\,\ 3: \quad\textbf{for} $other$ \textbf{in} $list_{TF}$ \textbf{do} \\
\STATE\,\ 4: \qquad\textbf{if} $node$ \textbf{in} $list_{TF}$ \textbf{or} $lat.in(other, node)$ \textbf{then}\\
\STATE\,\ 5: \qquad\quad \textbf{continue}\\
\STATE\,\ 6: \qquad \textbf{end if}\\
\STATE\,\ 7: \qquad \textbf{if} $other.LH$ \textbf{not in} $node.RH$ \textbf{then}\\
\STATE\,\ 8: \qquad\quad \textbf{continue}\\
\STATE\,\ 9: \qquad \textbf{end if} \\
\textbf{Field-Sensitive Analysis:}\\
\STATE10: \qquad $tainted\_id=trace\_tainted\_attr(node)$ \\
\STATE11: \qquad $tainted\_attr.append(tainted\_id)$ \\
\STATE12: \qquad \textbf{if} ($hasattr(node.RH,attr)$ \textbf{and} $node.RH.attr$ \textbf{not in} $tainted\_attr$) \textbf{then}\\
\STATE13: \qquad\quad \textbf{continue}\\
\STATE14: \qquad \textbf{end if} \\
\STATE15: \qquad $list_{TF}.append(node)$ \\
\STATE16: \quad \textbf{end for} \\
\STATE17: \textbf{end for} \\
\end{algorithmic}
\label{alg2}
\end{algorithm}

\textbf{Framework-Adapted Module.} To reduce the incidence of false positives in static taint analysis, we incorporate the dispatching rules of the LuCI framework into our validation process for potential vulnerabilities. In the MVC architecture framework, the controller module serves as the pivotal component of the web application request processing pipeline. It is responsible for receiving input data from web requests and directing this information to the appropriate model or view modules. LuCI follows the MVC paradigm, which suggests that external inputs are primarily derived from the controller module. Consequently, in taint detection, we implement enhanced input validation mechanisms in the \texttt{luci/controller} directory. This approach aims to determine whether the existing data flow can precipitate vulnerabilities, particularly focusing on data that might be maliciously injected via URLs and their associated parameters. For alarms triggered in other segments of the application, if the originating source is determined not to stem from the HTTP request interface, these instances are systematically excluded from consideration.

\subsection{Pruning with LLM Assistance}
Beyond the false positives that can be resolved through the dispatching rules of the framework, some false alarms persist and cannot be bypassed simply by establishing specific rules. Considering the recent popularity of LLMs for performing code analysis \cite{ZhangZiyin2023, CodaMosa2023}, we envision using LLM to assist our analysis in the pruning of the generated alarms. Fortunately, studies have shown the feasibility of this idea through small-scale experiments \cite{LiHaonan2023}. We choose to use Generative Pre-trained Transformer 4 (GPT-4) because GPT-4 outperforms existing large language models on a collection of NLP tasks and exceeds the vast majority of reported state-of-the-art systems (which often include task-specific fine-tuning)\cite{openai2023gpt4}.

\begin{figure}
\includegraphics[width=0.7\linewidth]{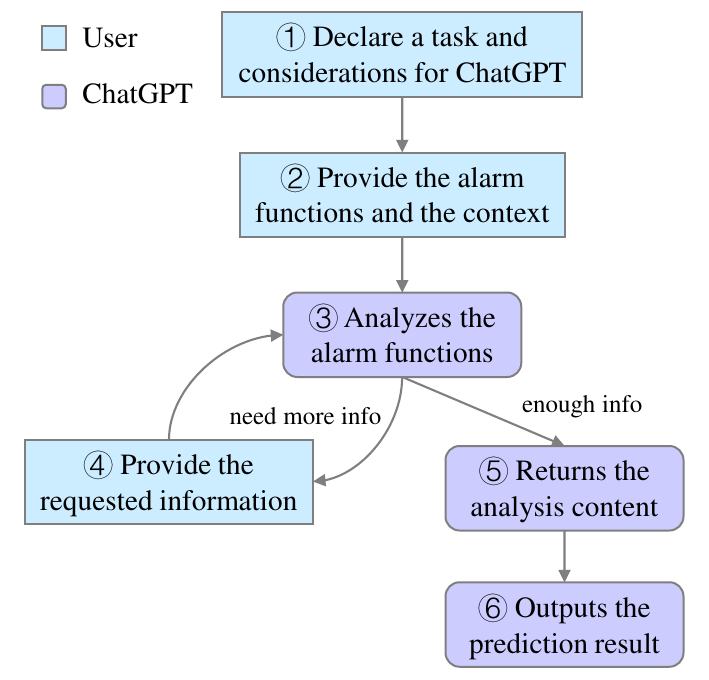}
\centering
\caption{\label{fig:5}Workflow of pruning with LLM assistance.}
\end{figure}

We summarize the causes of false alarms and then use the latest GPT-4 model to analyze the alarms by elaborately constructing problems to prune false alarms. This involves carefully crafting input prompts that allow the LLM to understand the context and intent behind the code, and then using its output to determine whether the alarm is a true positive or a false positive. Figure \ref{fig:5} shows the workflow of pruning with LLM assistance.

To illustrate this process, we show an example of conversation for function evaluation using GPT-4 on webpage\footnote{\href{https://chat.openai.com/share/4689d438-6867-490d-ae8c-9d963323f17e}{https://chat.openai.com/share/4689d438-6867-490d-ae8c-9d963323f17e}}. In this example, we provide the LLM with a code snippet and ask it to evaluate the function's behavior. The LLM responds with a detailed analysis of the code, highlighting potential issues and drawing conclusions about whether it is a true alarm.

\textbf{Integration of Pruning.} OpenAI has released an API for GPT-4 that allows developers and enterprises to integrate its advanced large-scale language model into their various applications and services. The pruning process is critical because it involves the analysis and filtration of data, identifying and eliminating false positives. This step essentially fine-tunes the selection procedure and streamlines the manual verification of vulnerabilities with alarms. We integrated the pruning process into the vulnerability detection system through the GPT-4 API and eventually saved the output of bug information in JavaScript Object Notation (JSON) format to automate the entire process.

\section{Implementation} \label{Implementation}
LuaTaint is implemented using approximately 12,000 lines of Python code. The syntax parsing module is constructed using the luaparser \cite{luaparser2023}, a Python-based Lua parser and AST builder. The control flow analyzer systematically traverses all the nodes based on their categories to establish a comprehensive CFG. The reaching definitions analyzer relies on reaching definitions worklist analysis and fixed-point algorithm, addressing data flow constraints through iterative processes. The framework-adapted module is designed to encapsulate the dispatching rules of various web frameworks, and it plays a pivotal role in vulnerability verification to mitigate false positives. The taint analyzer uses the constraint relationship between the nodes to conduct a backtracking analysis based on the arguments of the sink functions. During this process, the dispatching rules of the various web frameworks are combined. Using the GPT-4 API key, the pruning processor inputs the vulnerability-related functions reported in the above process into the content and outputs the predicted label of the potential vulnerability.

\section{Evaluation} \label{Evaluation}
We formulate several research questions to guide our assessment of the performance and efficiency of LuaTaint in identifying vulnerabilities.

\textbf{RQ1:} Is LuaTaint capable of detecting both manufactured vulnerabilities and those present in real-world IoT devices?

\textbf{RQ2:} How effective is the framework-adapted module within LuaTaint? How accurately can it prune false alarms using the LLM?

\textbf{RQ3:} In comparison with existing static analysis tools for Lua, how does LuaTaint perform in terms of performance on current tasks?

\textbf{RQ4:} What is the efficiency of LuaTaint? To what extent can the approximation operation reduce the overhead?

In this section, the answers to the aforementioned questions are obtained through several empirical investigations.

\subsection{Vulnerabilities Detection}
To investigate \textbf{RQ1}, we conducted experimental validation using manufactured vulnerabilities, real-world vulnerabilities, and relevant firmware.

\begin{table}
\caption{\label{tab:1}Results of Manufactured Vulnerabilities for Verification.}
\centering
\begin{tabular}{lccccc}
\toprule
\textbf{Type} & \textbf{CI} & \textbf{RCE} & \textbf{PAT} & \textbf{SQL} &  \textbf{Total}\\
\midrule
Number & 110 & 67 & 78 & 68 & 323\\
Recall Rate & 100.00\% & 91.04\% & 98.72\% & 100\% & 97.80\% \\
\bottomrule
\end{tabular}
\end{table}

\textbf{Manufactured Vulnerabilities.} We initially assessed the effectiveness and completeness of LuaTaint by examining its performance on manufactured vulnerabilities. For this purpose, we manually composed 323 distinct examples of known vulnerabilities, including CI, RCE, path traversal (PAT), and SQL injection. These examples were meticulously designed to encompass a variety of scenarios, incorporating diverse sensitive functions and modes of parameter propagation, to thoroughly demonstrate the wide range of potential security vulnerabilities. We evaluated the completeness of LuaTaint on the aforementioned dataset and the results are shown in Table \ref{tab:1}. There are some missing reports in RCE and PAT types, and the total recall rate on the manufactured vulnerability dataset is about 97.80\%.  We analyze LuaTaint's missing reports individually and find that most of the missing reports are due to the failure of tracking the tainted data flow due to extremely complicated function invokes. Therefore, we have further improved the performance of LuaTaint based on the current missing reports so that it can accurately identify these missing reports.

\begin{table}
\caption{\label{tab:2}Results of Existing CVE List for Verification.}
\centering
\begin{tabular}{lclll}
\toprule
\textbf{CVE-ID} & \textbf{Vendor} & \textbf{Series/Version} & \textbf{Type} & \textbf{Alarm} \\
\midrule
CVE-2017-16957 & TP-Link & WVR/WAR/ER/R & CI & True \\
CVE-2017-16958 & TP-Link & WVR/WAR/ER/R & CI & True \\
CVE-2017-16959 & TP-Link & WVR/WAR/ER/R & CI & True \\
CVE-2017-17757 & TP-Link & WVR/WAR/ER/R & CI & True \\
CVE-2018-11481 & TP-Link & IPC & RCE & False \\
CVE-2019-12272 & OpenWrt & LuCI 0.10 & CI & True \\
CVE-2019-19117 & Phicomm & K2 & CI & True \\
CVE-2021-28961 & OpenWrt & OpenWrt 19.07 & CI & True \\
CVE-2021-43162 & Ruijie & RG-EW & RCE & Encypto \\
CVE-2022-28373 & Verizon & ODU & RCE & True \\
CVE-2022-28374 & Verizon & ODU & RCE & True \\
\bottomrule
\end{tabular}
\end{table}

\textbf{Real-World Vulnerabilities.} After that, we evaluated the performance of LuaTaint on real-world vulnerabilities, particularly those associated with the LuCI framework in recent years. Table \ref{tab:2} presents a summary of these vulnerabilities, detailing their CVE-ID, the affected vendors, the device series or versions, and the types of vulnerabilities identified. These vulnerabilities were primarily found in the firmware of certain devices from specific vendors and in older versions of the OpenWrt distribution, with CI and RCE being the predominant types. We used LuaTaint to automate the detection of these vulnerabilities and conduct a thorough analysis of our results. It is noteworthy that Ruijie's firmware is encrypted, which obstructs the ability to unpack and analyze the source code. Additionally, the TP-Link IPC series, which is designed for visual security through smart cameras, includes a vulnerability (CVE-2018-11481) that allows RCE. This particular vulnerability arises from the lenient restrictions in the string matching process using \texttt{"\%p"}, in which LuaTaint is not identified as an alarm owing to the lack of an executable function and the resulting gap in the identification of hazardous data flows. However, LuaTaint successfully identifies the remaining vulnerabilities associated with the provided CVE IDs as alarms.

Afterward, we collected \textbf{2,447} firmware samples from 11 vendors, including 8devices, Avalon, GL.iNet, Fastcom, Linksys, NetGear, TP-Link, Xiaomi, X-Wrt, LibreMesh, and OpenWrt, for testing, as illustrated in Table \ref{tab:3}. We classified an issue that affects the same function and argument across various devices as a single alarm and verified the results of \textbf{145} tested firmware samples. LuaTaint detected \textbf{111} true alarms in these 145 firmware samples, and \textbf{102} of which were previously unknown. Most of these vulnerabilities are due to inadequate input filtering, leading to CI and RCE, but we also found five path traversal vulnerabilities. Based on the serious security implications, we have reported these defects to the respective device vendors. Eight of these bugs have been confirmed by the vendors as present in older stable firmware versions, while the status of the others is still under review. These findings demonstrate the effectiveness of LuaTaint in uncovering common vulnerabilities in the web configuration interfaces of IoT devices. 

\begin{table}
\caption{\label{tab:3}Datasets and Results of Firmware under Testing.}
\centering
\scalebox{0.81}{
\begin{threeparttable}
\begin{tabular}{llccccc}
\toprule
\textbf{Vendor} & \textbf{Device Type} & \textbf{TImg} & \textbf{Alarms} & \textbf{CImg} & \textbf{Bugs} & \textbf{LImg} \\
\midrule
8devices & System-on-Module & 9 & 2 & 9 & 2 & 8\\
Avalon & Miner & 18 & 4 & 9 & 3 & 9\\
GL.iNet & Gateway, Router & 87 & 5 & 18 & 4 & 4\\
Fastcom & Router, Others & 11 & 39 & 11 & 17 & 11\\
Linksys & Router & 2 & 24 & 2 & 24 & 2\\
NetGear & Router & 3 & 24 & 3 & 24 & 2\\
TP-Link & Router, Switch, Others & 29 & 67 & 10 & 33 & 10\\
Xiaomi & Router, Others & 21 & 1 & 21 & 0 & 0\\ 
OpenWrt & AccessPoint, Router, Others & 2225 & 6 & 20 & 4 & 15\\
LibreMesh & AccessPoint, Router, Others & 24 & 0 & 24 & 0 & 0\\
X-Wrt & AccessPoint, Router, Others & 18 & 0 & 18 & 0 & 0\\\midrule
\textbf{Total} &  & \textbf{2447} & \textbf{172} & \textbf{145} & \textbf{111} & \textbf{61}\\
\bottomrule
\end{tabular}
\begin{tablenotes}
\footnotesize
\item Note: \textbf{TImg} denotes the number of tested firmware images. \textbf{CImg} denotes the number of verified images. \textbf{Bugs} denotes the number of verified bugs. \textbf{LImg} denotes the number of verified leaky images.
\end{tablenotes}
\end{threeparttable}}
\end{table}

\textbf{Key Insights:} It is important to observe that the same vulnerability codes recur across various firmware series from identical vendors. Furthermore, instances of identical vulnerability codes are identified in the firmware of multiple disparate vendors. This occurrence may be attributed to a common practice among smart device vendors who, upon integrating OpenWrt into their systems, might overlook conducting thorough security evaluations of the respective versions. Consequently, vulnerabilities intrinsic to the original OpenWrt system continue to manifest in smart devices produced by prominent manufacturers. Based on these observations, LuaTaint can help developers identify these vulnerabilities early on, reducing their spread and enabling timely fixes. Ideally, these vulnerabilities should be detected before devices are put into use. Developers should adopt comprehensive strategies to improve the security of these devices, such as establishing a regular and effective update and patch management process to address known vulnerabilities quickly.

\lstset{
 basicstyle=\ttfamily\fontsize{6.2}{8}\selectfont,
 columns=fixed,
 numbers=left,                                        
 numberstyle=\tiny\color{gray},                       
 numbersep=5pt,                                       
 xleftmargin=1em,                                     
 aboveskip=3pt, 
 belowskip=-1em, 
 backgroundcolor=\color[RGB]{245,245,244},            
 keywordstyle=\color{blue},                 
 morekeywords={local,while,do,if,then,end,function,return,and,or,for,in},
 numberstyle=\footnotesize\color{darkgray},
 stringstyle=\rmfamily\slshape\color[RGB]{128,0,0},   
 emph={get_device_byif,tophy,check_section_available,add_br,io,popen},
 emphstyle=\color{magenta},
 showstringspaces=false,                              
 moredelim=[is][\color{cyan}]{\$}{\$},
 moredelim=[is][\color{SeaGreen}]{\&}{\&},
 label=lst:3,
}
\begin{lstlisting}[caption={Functions related to CVE-2017-16958 in WVR/WAR/ER/R \\series of TP-Link firmware.},captionpos=b]
---TL-WAR450Lv1.bin /usr/lib/lua/luci/controller/admin/bridge.lua
local function get_device_byif($iface$)
    local $mycmd$ = ". /lib/zone/zone_api.sh; zone_get_device_byif" 
    .. $iface$
    local ff = io.popen($cmd$, "r")
end
local function tophy($ifname$)
    local phy = {}
    for k,$v$ in ipairs($ifname$) do
        phy[#phy + 1] = get_device_byif($v$)
    end
    return phy
end
local function check_section_available($data$, op)
    local $new$ = { }
    $new$.&t_name& = $data$.&t_name&
    -- change the index of IF to name (total of 5 IFs)
    $new$.&ifname& = tophy($new$.&t_bindif&)
end
function add_br($http_form$)
    local $data$ = json.decode($http_form$.&data&)
    local $jdata$ = $data$.&params&
    local $input$ = $jdata$.&new&
    if not input or type($input$) ~= "table" then
        return false, err.ERR_COM_TABLE_ITEM_UCI_ADD
    end
    new = check_section_available($input$, "add")
end
\end{lstlisting}
\vspace{3mm}

\textbf{Case Study 1: Tracing Taint Flow.} 
To have a better understanding of LuaTaint taint tracking, we examine a specific case highlighting detected vulnerabilities, as shown in Listing \ref{lst:3}. In the TP-Link firmware, we identified the function \texttt{get\_device\_byif()} located at \texttt{"/squashfs-root /usr/lib/lua/luci/controller/admin/bridge.lua"}. This function exhibits vulnerability owing to the injection of the \texttt{"io.popen"} API. The popen parameter \texttt{cmd} concatenates the string \texttt{"./lib/zone/zone\_api.sh;zone\_get\_device \_byif"} with the \texttt{"iface"} variable. To understand the risk, we traced the call to the \texttt{get\_device\_byif()} function up to the HTTP parameter of the requested external variable. After tracing back to the injected API parameters,
the call sequence leading to the vulnerability is \texttt{add\_br(http \_form)$\xrightarrow{}$check\_section\_available(data,p)$\xrightarrow{}$tophy( ifname)$\xrightarrow{}$get\_device\_byif()}. The key point to note is that there is no external parameter string filtering in this sequence, leading to the classification of \texttt{get\_device\_byif()} as a command execution vulnerability, identified as CVE-2017-16958.

\textbf{Case Study 2: Command Injection.} 
Listing \ref{lst:4} shows a command injection vulnerability in a router firmware confirmed by vendors. This issue occurs after users successfully log in and access the \texttt{network.lua} file via the web manager, such as uhttpd. Here, users can add an interface using the \texttt{iface\_reconnect(iface)} function to the controller. Critically, the \texttt{iface\_reconnect} function does not validate or restrict the input parameters, leading to the direct use of the \texttt{luci.sys.call} function to execute shell commands, with the \texttt{"iface"} parameter taken directly from the HTTP post data. Therefore, attackers can exploit this vulnerability to execute harmful commands or to access confidential files by inserting specific system commands.

\lstset{
 basicstyle=\ttfamily\fontsize{7.0}{8}\selectfont,
 columns=fixed,
 numbers=left,                                        
 numberstyle=\tiny\color{gray},                       
 numbersep=5pt,                                       
 xleftmargin=1em,                                     
 aboveskip=3pt, 
 belowskip=-1em, 
 backgroundcolor=\color[RGB]{245,245,244},            
 keywordstyle=\color{blue},                 
 morekeywords={local,while,do,if,then,end,function,return,and,or,for,in},
 numberstyle=\footnotesize\color{darkgray},
 stringstyle=\rmfamily\slshape\color[RGB]{128,0,0},   
 emph={iface_reconnect, get_network, sys,call, },
 emphstyle=\color{magenta},
 showstringspaces=false,                              
 moredelim=[is][\color{cyan}]{\$}{\$},
 label=lst:4,
}
\vspace{-3mm}
\begin{figure}[H]
\begin{lstlisting}[caption={A command injection vulnerability in a router firmware.},captionpos=b]
-- usr/lib/lua/luci/controller/admin/network.lua
function iface_reconnect($iface$)
    local netmd = require "luci.model.network".init()
    local net = netmd:get_network($iface$)
    if net then
        luci.sys.call("env -i /sbin/ifup %q >/dev/null 2>
        /dev/null" % $iface$)
        luci.http.status(200, "Reconnected")
        return
    end
    luci.http.status(404, "No such interface")
end
\end{lstlisting}
\end{figure}
\vspace{-1mm}

\textbf{Case Study 3: Path Traversal.} 
Listing \ref{lst:5} illustrates a path traversal vulnerability found in a router firmware. This issue allows access to the \texttt{action\_del\_script()} function in \texttt{commands.lua} without requiring web log in authentication. User input is collected from a web form via \texttt{luci.http.formvalue("set")} and then assigned to the variable \texttt{name} using \texttt{luci.http.formvalue}. Crucially, this input does not undergo verification or restriction. Consequently, it is directly used by \texttt{os.remove} to delete files in the specified path of the input. Attackers can abuse this by submitting specific system file names, resulting in potential information leaks or system damage. In worst cases, this vulnerability can facilitate denial-of-service attacks.

\lstset{
 basicstyle=\ttfamily\fontsize{7.0}{8}\selectfont,
 columns=fixed,
 numbers=left,                                        
 numberstyle=\tiny\color{gray},                       
 numbersep=5pt,                                       
 xleftmargin=1em,                                     
 aboveskip=3pt, 
 belowskip=-1em, 
 backgroundcolor=\color[RGB]{245,245,244},            
 keywordstyle=\color{blue},                 
 morekeywords={local,while,do,if,then,end,function,return,and,or,for,in},
 numberstyle=\footnotesize\color{darkgray},
 stringstyle=\rmfamily\slshape\color[RGB]{128,0,0},   
 emph={action_del_script, formvalue, os, remove},
 emphstyle=\color{magenta},
 showstringspaces=false,                              
 moredelim=[is][\color{cyan}]{\$}{\$},
 label=lst:5,
}
\begin{lstlisting}[caption={A path traversal vulnerability in a router firmware.},captionpos=b]
-- /usr/lib/lua/luci/controller/commands.lua
function action_del_script()
    local name = luci.http.formvalue($"set"$)
    local rv ={}
    os.remove($name$)
    $rv$["name"] = $name$
    luci.http.prepare_content("application/json")
    luci.http.write_json($rv$)
end
\end{lstlisting}

\subsection{Evaluation of Accuracy Enhancements.}
We conducted three sets of experiments to verify the impact of the current enhancement designs on the effectiveness of vulnerability detection to answer \textbf{RQ2}.

\begin{table}
\caption{\label{tab:4}Results Before and After using Framework-Adapted Module.}
\centering
\scalebox{0.74}{
\begin{tabular}{llcccccc}
\toprule
\textbf{Vendor} & \textbf{Device} & ${Alarms_B}$ & ${FP_B}$ & ${FR_B}$ & ${Alarms_A}$ & ${FP_A}$ & ${FR_A}$ \\
\midrule
8devices & Carambola2 & 4 & 2 & 50.00\% & 11 & 9 & 81.82\%\\
8devices & Cherry & 2 & 0 & 0.00\% & 9 & 7 & 77.78\%\\
Avalon & Avalon2 & 15 & 13 & 86.67\% & 28 & 26 & 92.86\%\\
Avalon & Avalon761 & 16 & 13 & 81.25\% & 36 & 33 & 91.67\%\\
GL.iNet & MT300A & 18 & 14 & 77.78\% & 38 & 34 & 89.47\%\\
Linksys & EA8500-full & 28 & 4 & 14.29\% & 37 & 13 & 35.14\%\\
NetGear & WNDR3700-V4 & 26 & 3 & 11.54\% & 35 & 12 & 34.29\%\\
TP-Link & Archer C2600 & 22 & 21 & 95.45\% & 44 & 43 & 97.73\%\\
TP-Link & TL-WDR7300 & 4 & 2 & 50.00\% & 9 & 7 & 77.78\%\\
Xiaomi & R1350 & 9 & 9 & 100.00\% & 20 & 20 & 100.00\%\\
OpenWrt & Lede-17.01.6 & 15 & 13 & 86.67\% & 28 & 26 & 92.86\%\\\midrule
\textbf{Total} &  & 159 & 94 & \textbf{59.12\%} & 295 & 230 & 77.97\%\\
\bottomrule
\end{tabular}}
\end{table}

\textbf{Effectiveness of Framework-Adapted Module.} 
To assess the efficacy of the framework-adapted module, we conducted comparative experiments by examining the results of a taint propagation analysis with and without the framework-adapted module. Table \ref{tab:4} presents the statistical data on the number of firmware alarms and the number of false alarms under the two experimental conditions. We define the false rate (\emph{FR}) as the ratio of the number of false alarms to the total number of alarms. The three columns labeled with the subscript \emph{B} represent the results obtained before implementing the framework-adapted module, while those with the subscript \emph{A} demonstrate the results after its application. Considering the commonality of vulnerabilities between firmware samples from the same vendors, we chose firmware samples from various vendors to ensure a comprehensive and diverse dataset.

As observed, the framework-adapted module effectively eliminates most false alarms during the vulnerability verification process, thereby decreasing the overall false alarm rate from \textbf{77.97\%} to \textbf{59.12\%}. This significant reduction saves considerable time and resources in subsequent vulnerability verifications. In practical terms, we utilize the framework-adapted module to sift through tainted data flows that satisfy specific criteria. Although these data flows have sources, sinks, and paths between them, (a) in some cases, the functions triggering these alarms are called by other functions with constant parameter values for the tainted sources; and (b) in other cases, the function parameters are variables that cannot be manipulated through the interface, thereby preventing attackers from setting these parameter values. Consequently, actual remote attacks cannot be launched via these data flows, thus allowing us to filter them out using the features of the framework-adapted module before raising an alarm.

\begin{table}
\centering
\setlength{\abovecaptionskip}{0mm} 
\setlength{\belowcaptionskip}{-2mm}
\caption{\label{tab:5} Results Before and After LLM Pruning.}
\scalebox{0.7}{
\begin{tabular}{llcccccc}
\toprule
\textbf{Vendor} & \textbf{Device Series} & \textbf{Alarms} & \textbf{T/F} & \textbf{TP/FN} & \textbf{FP/TN} & $\boldsymbol{Pre_B}$ & $\boldsymbol{Pre_A}$\\
\midrule
\multirow{4}*{8devices} & Kinkan & 3 & 0/3 & 0/0 & 0/3 & 0.00\% & -\\
                        & Carambola2 & 4 & 2/2 & 2/0 & 0/2 & 50.00\% & 100.00\%\\
                        & Cherry & 2 & 2/0 & 2/0 & 0/0 & 100.00\% & 100.00\%\\
                        & Carambola3 & 4 & 2/2 & 2/0 & 0/2 & 50.00\% & 100.00\%\\
\rowcolor{Lavender} & Avalon2 & 15 & 2/13 & 2/0 & 2/11 & 13.33\% & 50.00\%\\
\rowcolor{Lavender} & Avalon4 & 15 & 2/13 & 2/0 & 1/12 & 13.33\% & 66.67\%\\
\rowcolor{Lavender} & Avalon741 & 15 & 2/13 & 2/0 & 1/12 & 13.33\% & 66.67\%\\
\rowcolor{Lavender} \multirow{-4}*{Avalon} & Avalon761 & 16 & 3/13 & 3/0 & 2/11 & 18.75\% & 60.00\%\\
\multirow{2}*{GL.iNet} & mt300a-2.265 & 18 & 4/14 & 4/0 & 1/13 & 22.22\% & 80.00\%\\
                       & mt300n-2.265 & 18 & 4/14 & 4/0 & 1/13 & 22.22\% & 80.00\%\\
\rowcolor{Lavender} & EA8500-full & 28 & 24/4 & 24/0 & 0/4 & 85.71\% & 100.00\%\\
\rowcolor{Lavender} \multirow{-2}*{Linksys} & EA8500 & 27 & 24/3 & 24/0 & 0/3 & 88.89\% & 100.00\%\\
\multirow{2}*{NetGear} & WNDR3700-V4-full & 28 & 24/4 & 24/0 & 0/4 & 85.71\% & 100.00\%\\
                       & WNDR3700-V4 & 26 & 23/3 & 23/0 & 0/3 & 88.46\% & 100.00\%\\
\rowcolor{Lavender} & Archer C2600 & 22 & 1/21 & 1/0 & 5/16 & 4.55\% & 16.67\%\\
\rowcolor{Lavender} \multirow{-2}*{TP-Link} & TL-WDR7300 & 4 & 2/2 & 2/0 & 0/2 & 50.00\% & 100.00\%\\
\multirow{2}*{Xiaomi} & R1350 & 9 & 0/9 & 0/0 & 1/8 & 0.00\% & 0.00\%\\
                      & R1cm & 2 & 0/2 & 0/0 & 0/2 & 0.00\% & -\\
\rowcolor{Lavender} & lede-17.01.6 & 15 & 2/13 & 2/0 & 1/12 & 13.33\% & 66.67\%\\
\rowcolor{Lavender} & openwrt-19.07.9 & 2 & 2/0 & 2/0 & 0/0 & 100.00\% & 100.00\%\\
\rowcolor{Lavender} \multirow{-3}*{OpenWrt} & openwrt-21.02.1 & 0 & 0/0 & 0/0 & 0/0 & - & -\\
\textbf{Total} &  & 273 & 125/148 & 125/0 & 15/133 & 45.79\% & \textbf{89.29\%}\\
\bottomrule
\end{tabular}}
\end{table}

\textbf{Accuracy of LLM Pruning.} 
In addition, we evaluated the performance of the GPT-4 API on a selected range of firmware from various vendors. Table \ref{tab:5} illustrates the statistics of the vulnerabilities detected by LuaTaint in some firmware samples. \emph{Alarms} refers to the number of alarms analyzed by LuaTaint before pruning, while \emph{T/F} denotes the count of true and false alarms post-manual verification. True positive (\emph{TP}), true negative (\emph{TN}), false positive (\emph{FP}), and false negative (\emph{FN}) denote the results of the LLM prediction label compared with those of the ground truth. To minimize random errors, each of these labels underwent five assessments, guaranteeing the highest level of confidence in the results. $Pre_B$ and $Pre_A$ denote the precision of the alarms before and after the pruning process, respectively. In these experiments, the assessment included not only false alarms but also true alarms. The experimental results indicate that GPT-4 demonstrates relatively stable performance in identifying true alarms, achieving a 100\% recall rate (total $TN=0$). Moreover, the precision in identifying false alarms was significantly enhanced from \textbf{45.79\%} to \textbf{89.29\%}, thereby effectively reducing numerous false alarms.

Furthermore, we analyzed the alarms for both successful and failed prunings by the LLM. The key to success or failure lies in whether the LLM can accurately analyze the data processing functions. Actually, the unpruned alarms all exhibited data flows from source to sink. Some of these data flows are relatively simple, while others involve quite complex data processing. These complex processing functions cannot be accurately identified by fixed rules, nor can they reliably determine whether sanitization has occurred, which motivated our use of the LLM for pruning. The LLM analyzes the processing functions within the data flow to some extent and determines whether the data has been sanitized during this process. Although it cannot guarantee completely accurate results, its extensive training on a large corpus of code significantly contributes to this process.

\begin{figure}
\setlength{\abovecaptionskip}{0mm}
\includegraphics[width=0.97\linewidth]{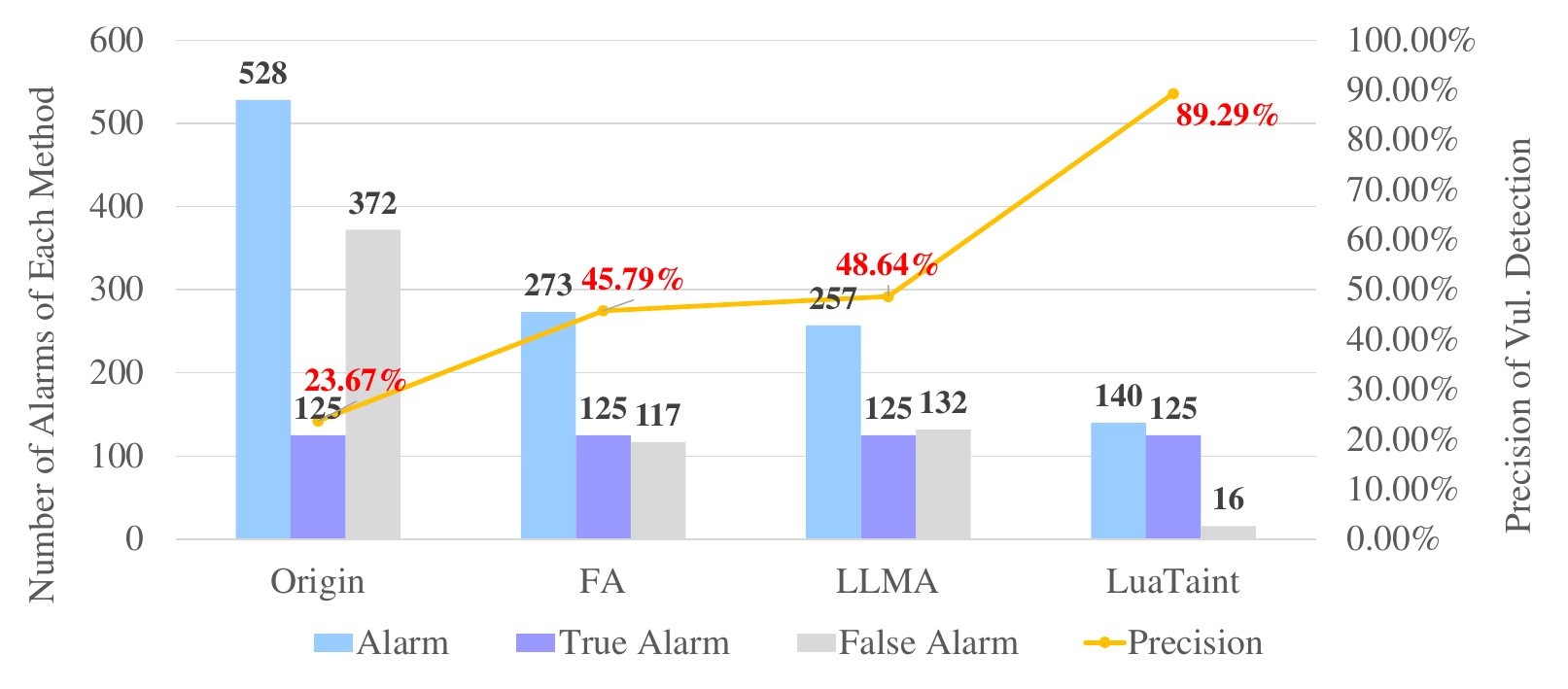}
\centering
\caption{\label{fig:6}Vulnerability Detection Precision results of ablation experiments.}
\end{figure}

\textbf{Ablation Studies.} To demonstrate the distinct effectiveness of our designed strategy for eliminating false positives, we conducted ablation studies on Framework-Adapted (FA) and LLM Assistance (LLMA) for the multi-vendor, multi-series firmware, as illustrated in Table \ref{tab:5}. The experimental results are depicted in Figure \ref{fig:6}. The framework-adapted module and the LLM pruning module improve the precision of vulnerability detection from \textbf{23.67\%} to \textbf{45.79\%} and \textbf{48.64\%}, respectively. Simultaneously, LLMA slightly outperforms FA. Moreover, when both enhancements are applied, there is a significant improvement in precision, increasing to \textbf{89.29\%}. These results demonstrate that integrating rule-based and data-driven approaches in the vulnerability detection workflow significantly boosts performance while minimizing false positives. 

Using a combination of these techniques allows our methodology to provide a thorough and nuanced analysis of firmware vulnerability. This holistic strategy is particularly vital in the intricate landscape of multi-vendor, multi-series firmware. This synergy increases the precision of detecting vulnerabilities and ensures the adaptability and scalability of our system. In conclusion, our innovative approach for reducing false positives, which combines framework adaptation with LLM pruning, has proven to be an effective solution for improving firmware security.

\subsection{Comparison}
For \textbf{RQ3}, we compared LuaTaint with various static analysis tools for Lua, including LuaCheck \cite{luacheck2018}, TscanCode \cite{TscanCode2022}, and Semgrep\cite{Semgrep2023}. We evaluated the effectiveness of these tools in analyzing Lua and the LuCI frameworks using the firmware sample dataset presented in Table \ref{tab:6}.

LuaCheck is mainly used to detect common problems in Lua code, such as syntax errors, code style problems, and unused variables. Although these checks contribute to improving the code quality and reducing errors, they lack a specific focus on the detection of security vulnerabilities. 
As shown in Table \ref{tab:6}, LuaCheck detected \textbf{25} code defects in the firmware. Upon individual inspection, we found that these defects were syntax errors caused by improper use of escape characters, which prevented luaparser from parsing the source code correctly. TscanCode is a static code analysis tool developed by Tencent that supports several programming languages, including C\#, C++, and Lua. TscanCode has a convenient GUI interface and an optional static analysis rule package. However, the current rules defined by TscanCode for Lua do not include taint analysis. We selected all the available rules for Lua in TscanCode and summarized \textbf{441} code defects with serious and critical impacts. The most common error types identified in these alarms were the use of undefined functions, uninitialized variables, and variable type errors.

\begin{table}
\caption{\label{tab:6}Comparison with the Lua Static Analysis Tools.}
\centering
\scalebox{0.73}{
\begin{tabular}{ll|c|c|cc|cc}
\toprule
\textbf{Vendor} & \textbf{Device Series} & \textbf{LuaCheck} & \textbf{TscanCode} & \multicolumn{2}{c|}{\textbf{Semgrep}} & \multicolumn{2}{c}{\textbf{LuaTaint}}\\
 &  & \textbf{Defects} & \textbf{Defects} & \textbf{Alarms} & \textbf{Vulns} & \textbf{Alarms} & \textbf{Vulns}\\ \midrule
8devices & Kinkan & 2 & 35 & 0 & 0 & 1 & 0\\
8devices & Carambola2 & 0 & 8 & 0 & 0 & 2 & 2\\
8devices & Cherry & 0 & 7 & 0 & 0 & 2 & 2\\
Avalon & Avalon2 & 1 & 26 & 3 & 0 & 4 & 2\\
Avalon & Avalon761 & 2 & 36 & 4 & 0 & 5 & 3\\
Fastcom & FER450 & 6 & 58 & 33 & 0 & 39 & 17\\
GL.iNet & MT300A & 1 & 31 & 3 & 0 & 5 & 4\\
Linksys & EA8500-full & 0 & 21 & 24 & 22 & 24 & 24\\
NetGear & WNDR3700-V4 & 0 & 22 & 23 & 21 & 23 & 23\\
TP-Link & Archer C2600 & 3 & 36 & 8 & 0 & 6 & 1\\
TP-Link & TL-WDR7300 & 0 & 36 & 4 & 0 & 2 & 2\\
TP-Link & TL-WAR450L & 6 & 66 & 36 & 0 & 59 & 30\\
Xiaomi & R1350 & 2 & 26 & 1 & 0 & 1 & 0\\
OpenWrt & lede-17.01.6 & 2 & 33 & 1 & 0 & 2 & 2\\ \midrule
\textbf{Total} &  & \textbf{25} & \textbf{441} & \textbf{140} & \textbf{43} & \textbf{175} & \textbf{112}\\
\bottomrule
\end{tabular}}
\end{table}

Additionally, the open-source static code analysis tool Semgrep also supports static analysis of Lua. It allows for fast code scanning of multiple languages and the creation of custom rules as needed. We migrated the trigger words for taint analysis from LuaTaint to Semgrep, customized a set of taint analysis rules for LuCI, and applied them to our current firmware sample set. As shown in Table \ref{tab:6}, the visual data indicates that while Semgrep did find some bugs in the firmware, its effectiveness was not as significant as LuaTaint's. Semgrep generated \textbf{140} alarms with \textbf{43} true positives, whereas LuaTaint produced \textbf{175} alarms with \textbf{112} true positives. These results demonstrate that Semgrep is less precise and comprehensive than LuaTaint in taint analysis for Lua and LuCI. We specifically analyzed Semgrep's results and identified its deficiencies: Firstly, Semgrep has limited ability to analyze taint propagation, and even a simple function encapsulation in the taint data flow can prevent it from tracing the flow accurately. Secondly, Semgrep's taint propagation analysis is restricted to using $\langle$\emph{sources, sinks, sanitizers}$\rangle$ rules, making it unable to detect dangerous data flows injected from URLs due to the non-fixed pattern of such data sources. In contrast, LuaTaint adapts to this with its framework-adapted module. Besides, we found that the vulnerabilities identified by Semgrep had relatively simple triggering mechanisms.

The results of the comparative experiments underscore the unique strengths and irreplaceability of LuaTaint for static analysis of Lua source code and the LuCI frameworks. The specialized capabilities of LuaTaint are essential for achieving more precise and comprehensive taint analysis in these contexts.

\subsection{Overhead}
We evaluated the operational cost of a segment of firmware, considering both the runtime and memory consumption of the process, to validate the effectiveness of the optimization of the approximate operation for \textbf{RQ4}. The pruning process is not included here to evade contingencies in the process of obtaining GPT-4 API replies (e.g., token limit, etc.). To avoid accidental errors, each sample is executed ten times, and the averages are calculated. The statistical results are presented in Table \ref{tab:7}. In this table, \emph{\#Fun} and \emph{\#Line} denote the total number of functions and lines of code present in LuCI, respectively, while \emph{Time} and \emph{Mem} denote the duration and memory expenditure required by the current firmware to complete an analysis. The subscripts \emph{B} and \emph{A} are utilized to distinguish between the states before and after the application of the approximation operation, respectively.

\begin{table}
\caption{\label{tab:7}Overhead Statistics for Some Firmware Samples on LuaTaint.}
\centering
\scalebox{0.7}{
\begin{tabular}{llcccccc}
\toprule
\textbf{Vendor} & \textbf{Device} & \textbf{\emph{\#Fun}} & \textbf{\emph{\#Line}} &  $\boldsymbol{Time_B}$ & $\boldsymbol{Mem_B}$ & $\boldsymbol{Time_A}$ & $\boldsymbol{Mem_A}$\\
\midrule
8devices & Carambola2 & 287 & 5875 & 14.53s & 66.32 MB & 13.47s & 54.91 MB\\
8devices & Cherry & 266 & 5836 & 4.22s & 44.77 MB & 3.65s & 45.97 MB\\
Avalon & Avalon2 & 989 & 20801 & 247.79s & 283.34 MB & 37.22s & 104.77 MB\\
Avalon & Avalon761 & 1046 & 20069 & 323.75s & 323.80 MB & 23.82s & 99.18 MB\\
GL.iNet & MT300A & 1109 & 21335 & 442.86s & 404.95 MB & 26.42s & 106.09 MB\\
Linksys & EA8500-full & 1294 & 22653 & 233.69s & 192.20 MB & 38.32s & 109.90 MB\\
NetGear & WNDR3700-V4 & 1244 & 21305 & 157.09s & 174.12 MB & 33.14s & 110.01 MB\\
TP-Link & Archer C2600 & 1883 & 39427 & 718.77s & 617.67 MB & 493.18s & 431.09 MB\\
TP-Link & TL-WDR7300 & 967 & 10369 & 322.07s & 316.05 MB & 174.99s & 161.79 MB\\
Xiaomi & R1350 & 1457 & 30056 & 167.92s & 276.78 MB & 66.80s & 153.65 MB\\
OpenWrt & lede-17.01.6 & 992 & 19533 & 372.93s & 379.90 MB & 22.27s & 97.11 MB\\\midrule
\bf{Average} &  & 1049 & 19751 & 273.24s & 279.99 MB & 84.84s & 134.04 MB\\
\bottomrule
\end{tabular}}
\end{table} 

From the firmware analysis of the two devices from the vendor \emph{8devices}, it is evident that when the LuCI program is relatively small, our approximation operation does not have a significant effect. However, as the size of the LuCI program increases, encompassing more functions, our approximation operation effectively reduces the operational costs of vulnerability detection, significantly decreasing both runtime and memory usage. The firmware tested in this experiment exhibited an average reduction in time and memory overhead of \textbf{68.95\%} and \textbf{52.13\%}, respectively. This demonstrates that our approach can effectively reduce computational complexity when handling larger and more complex firmware systems, making it a valuable tool for practical applications.

To provide a comprehensive understanding of the system's overhead, we enumerated the number of user prompts and tokens consumed during the pruning process. We have made some efforts to reduce energy costs associated with GPT-4 operations. Specifically, since identical vulnerability codes often recur across various firmware series from the vendors, we prune the same function only once. During the pruning process, a total of \textbf{162$\times$5} prompts were used, consuming approximately \textbf{1920k} tokens. These tokens are further divided into prompt tokens and response tokens, with approximately 1700k and 220k tokens respectively. These statistics reflect the energy consumption of GPT-4 in this process to a certain extent.

\section{Discussion} \label{Discussion}
Although LuaTaint is effective in discovering vulnerabilities in the web configuration interface of IoT devices, several shortcomings highlight them as key opportunities for future work.

\textbf{Lack of Front-end Information.} In this study, LuaTaint focuses more on the back-end of the page handler to detect vulnerabilities in web applications and lacks analysis of front-end web information. The data interaction between the front-end and back-end is often idiosyncratic and complex, and it is difficult to summarize a certain pattern in the static analysis of numerous objects. To expand our research, we plan to apply some user input information from the front-end of the page to further confirm the vulnerability so that LuaTaint can effectively detect XSS or certain types of sensitive information leakage vulnerabilities.

\textbf{Limitations of Analytical Methods.} We did not prioritize efficiency in this study because our goal was to discover more unknown vulnerabilities. However, the data flow analysis of a large project incurs a significant computational overhead. Other more efficient data flow algorithms can be used to improve data flow analysis in subsequent studies. Some improved data flow analysis algorithms, such as sparse data flow analysis\cite{shi2018pinpoint}, which focuses only on the variables that may affect the results of the analysis, reduces the amount of computation, and improves the efficiency of the analysis; incremental data flow analysis \cite{shao2010optimizing} utilizes the information from the previous analysis results and calculates only the parts of the program that have changed, thus reducing the cost of reanalyzing the entire program. Furthermore, we will continue to leverage open-source pre-trained LLMs and fine-tune them based on current tasks to improve the pruning performance of LLMs.

Future improvements could be made to increase the precision and efficiency of vulnerability identification in complex firmware ecosystems by incorporating additional data sources and refining analytical methods.

\section{Related Work} \label{RelatedWork}
We summarize the static, dynamic, and other analysis techniques for IoT firmware vulnerability detection in recent years, as well as security analysis methods for web configuration interfaces.

\textbf{Firmware Security Analysis.} Static and dynamic analyses have been used extensively in research to detect firmware vulnerabilities \cite{pistoia2007survey, li2010comparative}. Firmalice is a notable binary analysis framework that uses symbolic execution and program slicing to identify authentication bypass vulnerabilities in binary firmware \cite{shoshitaishvili2015firmalice}. Dtaint is a static binary taint analysis system \cite{cheng2018dtaint}, while Genius employs static code similarity-based analysis, transforming CFGs into numeric feature vectors to identify vulnerabilities \cite{feng2016scalable}. Other notable systems include KARONTE, a multi-binary static vulnerability detection system \cite{redini2020karonte}, and SATC, which uses shared keywords for taint analysis \cite{chen2021sharing}. SRFuzzer \cite{zhang2019srfuzzer} and IoTFuzzer \cite{chen2018iotfuzzer} are notable frameworks for fuzzy testing in router web servers and IoT devices, respectively. Tools, such as IoTHunter \cite{khandait2021iothunter}, Firm-AFL \cite{zheng2019firm}, and RPFuzzer \cite{wang2013rpfuzzer} further augment this approach. Representative tools for dynamic taint analysis include TaintCheck \cite{newsome2005dynamic}, Flayer, and BitBlaze. These dynamic techniques are instrumental in identifying vulnerabilities that may not be apparent through static analysis alone.

\textbf{Web Interface Security Analysis.} There are also some corresponding studies for web applications beyond firmware interfaces \cite{agosta2012automated,zheng2013path, gupta2014static,tripp2009taj}. For example, the TAJ \cite{tripp2009taj} tool employs hybrid slicing combined with object-sensitive alias analysis for taint analysis in Java web applications. Costin et al. \cite{costin2016automated} proposed a scalable and fully automated dynamic analysis framework to discover web interface-related vulnerabilities in firmware. Various issues must be addressed when using these tools because they are not specialized for vulnerabilities in IoT web interfaces or integrated into automated frameworks. FIRMADYNE \cite{chen2016towards} allows web penetration testing of accessible web interfaces over a local network. However, it has a low simulation of network reachability and web service availability. WMIFuzzer \cite{wang2019discovering} focuses on the web interface by enforcing user interface (UI) automation to drive the web interface to generate initial seed messages automatically. It also proposes the weighted message parse tree to guide mutations and uses fuzzy testing technology to detect vulnerabilities. The system has difficulty efficiently recognizing interactive features on the varied and complex web interfaces of IoT devices.

In addition, in recent years, significant studies have been conducted using natural language processing (NLP) techniques and LLMs to address issues related to vulnerability detection \cite{ziems2021security, thapa2022transformer, ye2023cp, ye2023tram}, which have yielded positive results. LuaTaint studied in this paper focuses on web configuration interfaces in firmware for a wide range of interactive intelligent devices, combining static taint analysis with LLM. It effectively detected real vulnerabilities in numerous IoT devices.

\section{Conclusion} \label{Conclusion}
We highlight a vital research direction for the security analysis for web configuration interfaces of IoT devices and present LuaTaint, an automated system designed to detect vulnerabilities in these interfaces. The system starts by parsing the web configuration interface code using Lua AST and creating CFGs. It then analyzes the data flow using reaching definitions and fixed-point algorithms to identify the constraints. Additionally, the system incorporates a taint analysis module tailored to the framework, along with LLM pinpointing and reporting vulnerabilities. The experimental results show that incorporating framework dispatching rules into taint analysis effectively reduces false alarms and that LLM can be used for efficient pruning. The combination of the two methods can achieve a vulnerability detection precision as high as 89.29\% in practical applications. LuaTaint successfully identified 111 vulnerabilities among 145 firmware samples, demonstrating its effectiveness in widespread and accurate bug detection in firmware web configuration interfaces. Although this study has made some progress in analyzing web configuration interface vulnerabilities of IoT devices, there is still room for improvement. Future work could further expand the research to consider more types of devices and vulnerability scenarios. In addition, more intelligent and automated vulnerability detection methods can be explored to improve detection efficiency and precision. These efforts will help strengthen the security of IoT devices and protect the security of user data and privacy.

\section*{Acknowledgments}
This work was partly supported by NSFC under Grant No. 62302443, the Fellowship of China National Postdoctoral Program for Innovative Talents (BX20230307), the Fundamental Research Funds for the Central Universities (Zhejiang University NGICS Platform). 


\end{document}